\documentclass[aip,jcp,reprint]{revtex4-1}

\usepackage{graphicx}  
\usepackage{dcolumn}   
\usepackage{bm}        
\usepackage{amssymb}   
\usepackage{url}
\usepackage{natbib}
\usepackage{color}
\usepackage[T1]{fontenc}
\usepackage[utf8]{inputenc}
\usepackage[normalem]{ulem}

\usepackage{graphicx}
\usepackage{amsmath}

\begin{document}

\title{
  A comparison between quantum chemistry and quantum Monte Carlo techniques
  for the adsorption of water on the (001) LiH surface}

\author{Theodoros~Tsatsoulis}
\affiliation{Max Planck Institute for Solid State Research, Heisenbergstra{\ss}e 1, D-70569 Stuttgart, Germany}
\author{Felix~Hummel}
\affiliation{Max Planck Institute for Solid State Research, Heisenbergstra{\ss}e 1, D-70569 Stuttgart, Germany}
\author{Denis~Usvyat}
\affiliation{Institut f{\"u}r Chemie, Humboldt-Universit{\"a}t zu Berlin, Brook-Taylor-Str. 2, D-12489 Berlin, Germany}
\author{Martin~Sch\"utz}
\affiliation{Institut f{\"u}r Chemie, Humboldt-Universit{\"a}t zu Berlin, Brook-Taylor-Str. 2, D-12489 Berlin, Germany}
\author{George~H.~Booth}
\affiliation{Department of Physics, King's College London, Strand, London, WC2R 2LS, United Kingdom}
\author{Simon~S.~Binnie}
\affiliation{London Centre for Nanotechnology, Gordon St., London WC1H 0AH, United Kingdom}
\affiliation{Thomas Young Centre, University College London, London WC1H 0AH, United Kingdom}
\affiliation{Department of Physics and Astronomy, University College London, London WC1E 6BT, United Kingdom}
\author{Michael~J.~Gillan}
\affiliation{London Centre for Nanotechnology, Gordon St., London WC1H 0AH, United Kingdom}
\affiliation{Thomas Young Centre, University College London, London WC1H 0AH, United Kingdom}
\affiliation{Department of Physics and Astronomy, University College London, London WC1E 6BT, United Kingdom}
\author{Dario~Alf\`e}
\affiliation{London Centre for Nanotechnology, Gordon St., London WC1H 0AH, United Kingdom}
\affiliation{Thomas Young Centre, University College London, London WC1H 0AH, United Kingdom}
\affiliation{Department of Physics and Astronomy, University College London, London WC1E 6BT, United Kingdom}
\affiliation{Department of Earth Sciences, University College London, London WC1E 6BT, United Kingdom}
\author{Angelos~Michaelides}
\affiliation{London Centre for Nanotechnology, Gordon St., London WC1H 0AH, United Kingdom}
\affiliation{Thomas Young Centre, University College London, London WC1H 0AH, United Kingdom}
\affiliation{Department of Physics and Astronomy, University College London, London WC1E 6BT, United Kingdom}
\author{Andreas~Gr\"uneis}
\email{a.grueneis@fkf.mpg.de}
\affiliation{Max Planck Institute for Solid State Research, Heisenbergstra{\ss}e 1, D-70569 Stuttgart, Germany}
\affiliation{Department Chemie, Technische Universit{\"a}t M{\"u}nchen, Lichtenbergstra{\ss}e 4, D-85747 Garching, Germany}

\date{\today}

\begin{abstract}
We present a comprehensive benchmark study of the adsorption energy of a single water molecule on the
(001) LiH surface using periodic coupled cluster and quantum Monte Carlo theories.
We benchmark and compare different implementations of quantum chemical wave function based theories in order to
verify the reliability of the predicted adsorption energies and the employed approximations.
Furthermore we compare the predicted adsorption energies to those obtained
employing widely-used van der Waals density-functionals.
Our findings show that quantum chemical approaches are becoming a robust
and reliable tool for condensed phase electronic structure calculations,
providing an additional tool that can also help in potentially improving currently
available van der Waals density-functionals.
\end{abstract}

\maketitle

\section{Introduction}

Kohn--Sham density-functional theory (DFT) is the standard
approach for the first-principles description of electronic properties
in computational material science and surface chemistry.
However, it is becoming clear
that the limitations of the employed exchange-correlation (XC) functionals
to balance off the numerous competing physical effects give rise to deficiencies in
the predictive ability of the approach, generally without any systematic manner to improve upon it.
One class of widely studied problems where this is particularly true is the case of
molecular adsorption on periodic surfaces.
Competing physical effects as well as poorly treated long-range dispersion contributions result in predicted
adsorption energies and sites varying strongly with the employed XC
functional (see \emph{e.g.} Refs.~\onlinecite{feibelman2001,li2008,schimka2010,karalti2012,alhamdani2015}).
This indicates fundamental shortcomings in many semi-local functionals that
are difficult to remedy.
Long-range dispersive interactions can be accounted for by the
addition of pairwise interatomic $C_{6} R^{-6}$
terms to the DFT energy, or by non-local functionals~\cite{tkatchenko2010,klimes2012,grimme2016}.
In this work we will refer to both the van der Waals corrected and van der Waals inclusive DFT
methods as van der Waals density-functionals.
Theoretically these corrections can be well justified and derived using quantum Drude
oscillators that serve as a qualitatively correct model for electrical response
properties between molecules and insulating solids.
However, most van der Waals corrections
also require the introduction of some adjustable parameters such as the
cutoff function and cutoff radius at short distances $R$ in order to remove the
attractive singularity from the $C_{6} R^{-6}$ terms.
These parameters can be obtained by optimizing the accuracy of the dispersion corrected
functionals for the description of molecular interaction energies in a given test set.


In this work, we consider an {\it ab-initio} description of
the true many-body wave function for a molecular adsorption problem.
Two contrasting yet complementary approaches
which we consider here, are
those from the field of quantum chemical Fock-space expansions of the wave function~\cite{bartlett2007},
and a stochastic representation from the Diffusion Monte Carlo (DMC) technique~\cite{foulkes2001}.
These wave function based approaches offer a thorough description of quantum
many-body effects through a direct treatment of electronic correlation. Such
approaches can supplement density-functional-based methods with accurate results.

DMC is a real-space quantum Monte Carlo (QMC) method, where the real-space
configurations of all $N$-electrons are sampled stochastically. This stochastic
distribution of electrons is evolved towards a sampling of the ground-state distribution of
electrons via an imaginary-time propagator, which exponentially filters out the higher-lying
eigenfunctions of the Hamiltonian from the distribution. This sampling would be exact if it were not
for the `Fermion sign problem', where the sampling collapses to the lower-energy symmetric
distribution of an $N$-particle Bosonic distribution. To avoid this, constraints are imposed
whereby the correct antisymmetry is maintained by imposing a hard nodal surface for the sampling
which enforces the sign of the sampled configurations. While this alleviates the Fermion sign problem,
it introduces a systematic and variational error due to this nodal surface, which in practical applications
is generally taken to be the nodal surface of a single Slater determinant. This represents the leading
error of a DMC calculation, but it benefits from a number of appealing properties which contrast
with quantum chemical methods, such as a very minor dependence on basis set, as well as a low-scaling
with respect to system size.
DMC techniques are increasingly used to understand molecular adsorption at
periodic surfaces~\cite{ma2011,karalti2012,alhamdani2015,wu2015}.

Quantum chemical methods constitute a hierarchy which
starting from the one-particle Hartree--Fock (HF) approximation, allows for a
systematic treatment of the quantum many-body effects.
The simplest form of such correlated methods is the second-order
M{\o}ller--Plesset perturbation theory (MP2).
Although MP2 theory provides a fair compromise between efficiency and
accuracy, certain effects are not captured accurately enough or at all
(\emph{e.g.} three-body dispersion interactions). For systems where such
effects are essential, the accuracy of the MP2 treatment is rather
modest. For instance, MP2 is known to notoriously overestimate dispersion
driven interactions in strongly polarizable systems~\cite{hesselmann2008,grueneis2010,sansone2016}. While many-body
perturbation theory offers a finite-order
approximation to electronic correlation, coupled-cluster theory provides a
compelling framework of infinite-order approximations in the form
of an exponential of cluster
operators.
The coupled-cluster singles and doubles (CCSD) method where the triples are treated in a
perturbative way, termed as CCSD(T), achieves chemical accuracy in the description of many molecular
properties and is often referred to as the gold standard method~\cite{bartlett2007}.
In recent years, quantum chemical wave function based methods have been
increasingly applied to periodic systems with the aim
of transferring their proven chemical accuracy in molecular systems
to the solid state~\cite{hirata2004,pisani2008,marsman2009,paier2009,binnie2010,voloshina2011,usvyat2011,booth2013,kubas2016,Yang2014,schwerdtfeger2016}.
However, the computational cost of quantum chemical wave function based methods
is a major obstacle for their application to extended systems.
The canonical formulation of MP2 theory
scales as $\mathcal{O}(N^5)$, where $N$ is a measure of the system
size, whereas CCSD theory
scales as $\mathcal{O}(N^6)$, and CCSD(T) as $\mathcal{O}(N^7)$.

This adverse scaling can in part be attributed to the use of canonical one-electron Bloch orbitals.
While canonical orbitals form a convenient basis for correlated
calculations since the Fock matrix is then diagonal, they are intrinsically
delocalized, rendering it difficult to build in the local character of electronic correlation.
In contrast, local correlation schemes~\cite{pulay1986,saebo1993} exploit the
fact that two-point correlations rapidly decay with distance in
insulating systems, by restricting excitations to spatially confined regions within
localized orbitals. It is possible to therefore reduce the scaling of the canonical
quantum chemical methods, in some cases to an asymptotic linear scaling~\cite{schuetz1999,schutz2000a}.
Several different local approximations exist, and represent a highly active field of research.
The method of increments relies on a similar local decomposition of the energy contribution, and
has been applied successfully to covalent large band-gap semiconductors, van der
Waals bonded rare-gas or molecular crystals, and molecular adsorption on
surfaces~\cite{stoll1992a,stoll1992b,stoll1992c,paulus2006,paulus2009,mueller2008,muller2013,hammerschmidt2012,Yang2014,weber2016}.


In this work, we will consider both local and canonical MP2 approaches in
similar basis sets, as well as comparing to both higher-level canonical coupled-cluster
and the contrasting DMC technique for the challenging problem of molecular
adsorption on a periodic surface.
Canonical CCSD theory will be explored
within the projector-augmented-wave (PAW) framework, using a plane-wave basis.
CCSD(T) theory will be applied in the form of corrections to MP2 with small
supercells and basis sets or using finite-clusters.
We assess the accuracy of these quantum
chemical schemes against the DMC results for water adsorption on the prototypical
ionic surface of lithium hydride (LiH). LiH has served as an important benchmark
system for several quantum-chemical methods~\cite{nolan2009,marsman2009,usvyat2011,stoll2012,vandevondele2012,booth2013,grueneis2015a} and
water adsorption on the (001) LiH surface can, in turn, be a benchmark
system for the interaction of molecules with surfaces. The relatively small
number of electrons involved allows for an in-depth comparison of different
post-mean-field methods.

Details about the structure of the system under consideration are
given in Sec.~\ref{sec:h2olih}. Computational details are presented in
Sec.~\ref{sec:computationalvasp}, Sec.~\ref{sec:computationalcrystal}, and
Sec.~\ref{sec:computationaldmc} for plane-wave, Gaussian basis, and DMC
calculations, respectively. Sec.~\ref{sec:results} summarizes all the results
obtained from different methods. Finally, we conclude the paper in
Sec.~\ref{sec:conclusions}.

\section{Computational Details}

\subsection{H$_2$O on LiH Geometry} \label{sec:h2olih}

\begin{figure}
 \includegraphics[width=7.0cm]{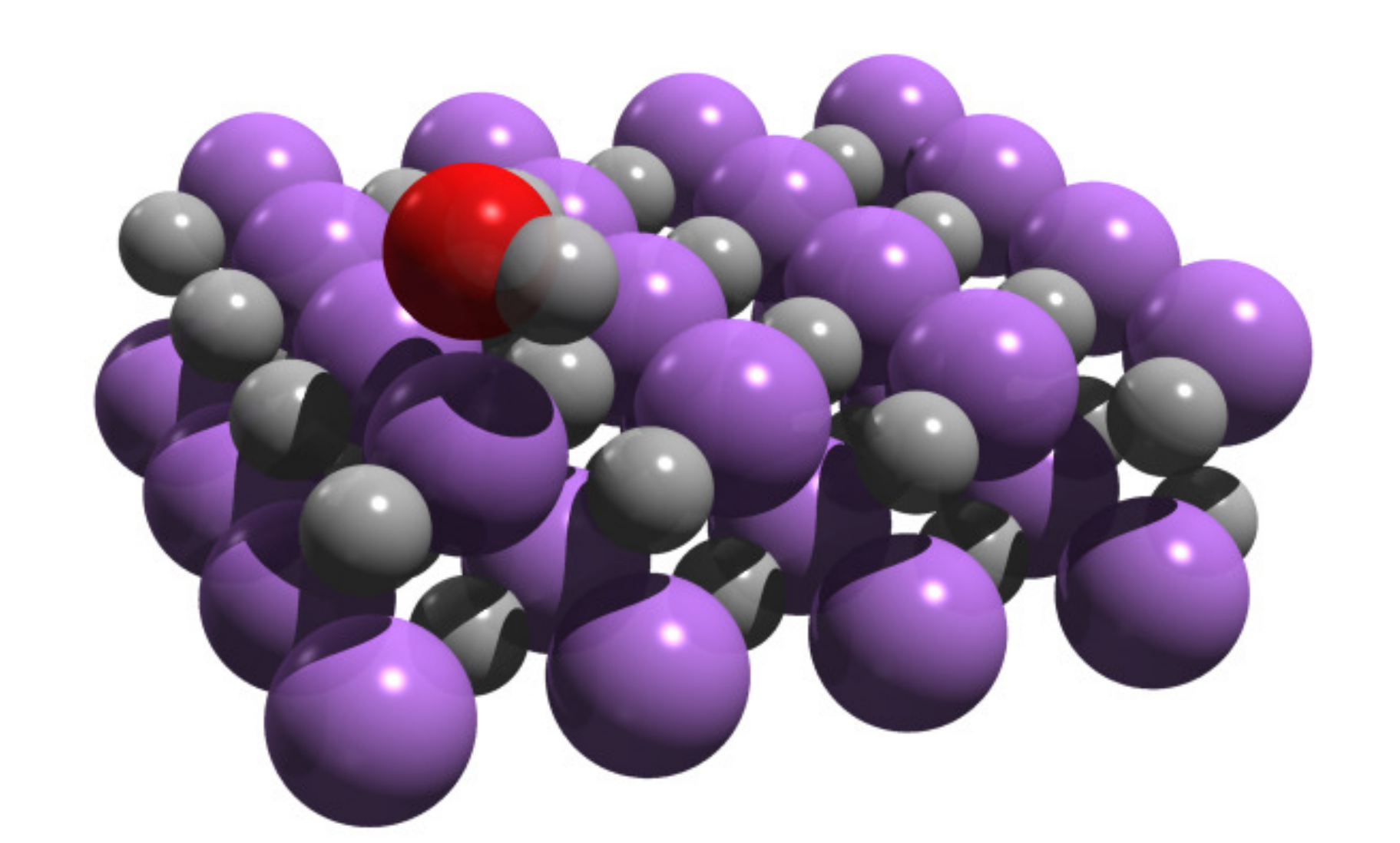}
  \caption{The adsorption geometry of water on a
    two-layer slab with 64 atoms per cell, representing the (001) LiH surface. The oxygen--surface distance is 2.15~\AA, while the water
    molecule almost retains its equilibrium structure. The geometry was
    optimized using the PBE functional.}
  \label{fig:adsorption}
\end{figure}

The aim of this work is to compare different high-level theories
for the calculation of the adsorption energy of a single water molecule
on the (001) LiH surface, keeping the atomic structure of the surface fixed.
The adsorption energy is defined as the
difference in energy between the non-interacting fragments
(water and the LiH surface) and the interacting system (water molecule on
LiH),
\begin{equation} \label{eq:adsorption}
  E_{\text{ads}} = E_{\text{H$_{2}$O}} + E_{\text{LiH}} -
    E_{\text{H$_{2}$O+LiH}}.
\end{equation}
An alternative definition for the adsorption energy is the difference between the
energy of the system with the water molecule at its equilibrium position on
the surface, and that of the system in which the water molecule has been
displaced vertically by 10~\AA. In both definitions the molecular structure of
the water molecule has been kept the same.
The latter definition is used for the DMC
calculations since it allows to maximize the possible cancellation of errors~\cite{zen16}.
We stress that since we are primarily interested in benchmarking different
electronic-structure methods, zero-point energy contributions or finite
temperature effects are neglected. The structure of the surface with the
adsorbed molecule has been obtained in the following manner. The Li and H
atoms have been kept fixed to their pristine lattice sites with a lattice
constant of $a = 4.084$ {\AA}, consistent with previous studies of the LiH
crystal~\cite{nolan2009,paier2009,binnie2010}. This has the advantage of
keeping the geometry consistent when supercells or fragments of different sizes
are used in quantum chemical and DMC calculations. The water molecule
was relaxed on the LiH (001) surface using the Perdew--Burke--Ernzerhof (PBE)
XC functional~\cite{perdew1996} and a two-layer slab with the $4\times4$ surface supercell.
For these calculations the {\sc vasp} code has been employed~\cite{kresse1999}.
A vacuum gap of 20.5 {\AA} has been employed to ensure that the surface slab
does not interact with its periodic image.
The relaxed geometry of the water molecule
adsorbed on the LiH surface is shown in Fig.~\ref{fig:adsorption}.
The DMC adsorption energy curve obtained by varying the distance between the
molecule and the surface, agrees well with the oxygen--surface distance of the
PBE functional (2.15~\AA)~\cite{binnie2011}. The structure of Fig.~\ref{fig:adsorption}
is given in the supplementary material.
This geometry is used throughout the paper for all density-functional and
correlated calculations. The convergence of the adsorption energy with the number
of layers in the slab is explored in Sec.~\ref{sec:layers}.


\subsection{Plane-Wave Basis Set Calculations} \label{sec:computationalvasp}

The calculations using a plane-wave basis set presented in this work
have been performed using the {\sc vasp} code
employing the PAW method alongside with the $\Gamma$-point approximation to
sample the first Brillouin zone.
The kinetic energy cutoff
that determines the size of the plane-wave basis set expansion of the
one-particle states was set to 500~eV.
There are numerous density-functionals that could be considered, of which we
have only chosen a small selection.
Thus, we assess the accuracy of one of the most widely-used functionals, the
PBE functional, as well as of several van der Waals functionals.
Specifically, dispersion corrections were taken into account following the
approach of Grimme~\textit{et al.}~\cite{grimme2010}, the method of Tkatchenko
and Scheffler~\cite{tkatchenko2009}, and the vdW-DF method proposed by
Dion~\textit{et al.}~\cite{dion2004,perez2009,klimes2010,klimes2011}, as
implemented in {\sc vasp}. In the former schemes a correction is added
to the DFT total energy after the self-consistent-field (SCF) cycle is
converged, whereas the latter scheme is a non-local correlation functional
that approximately accounts for dispersion interactions. 
In all calculations all electronic states of the H and Li atoms were treated
as valence states, whereas the $1s^2$ states of the O atom were kept
frozen. Supercells of different sizes were used to model the LiH surface,
containing 32, 64, and 128 atoms.

In the current paper we employ pseudized Gaussian-type orbitals (PGTOs)
expanded in a plane-wave basis set to span the virtual orbital manifold
necessary for the quantum chemical MP2 and coupled-cluster methods. The space of the
occupied orbitals from the HF calculation is projected out from the PGTOs,
ensuring that they solely span the virtual space. The rediagonalization of the
Fock matrix in this newly constructed virtual space allows for a canonical
formulation of quantum chemical techniques. This enables considerably fewer states to
be involved in many-body calculations~\cite{booth2016}. The method to obtain
PGTOs invokes a pseudization procedure of the sharply peaked Gaussian basis
sets, which follows the work of Kresse~\textit{et al.}~\cite{kresse1994}.
A more detailed explanation of PGTOs and their application to periodic systems
is given in Ref.~\onlinecite{booth2016}.
PGTOs allow for a controllable and reliable extrapolation of the adsorption energies
to complete basis set limit results.
For the present calculations Dunning's contracted aug-cc-pVDZ (AVDZ), aug-cc-pVTZ (AVTZ),
and aug-cc-pVQZ without $g$ functions (AVQZ--$g$) basis
sets~\cite{dunning1989,feller1996} were pseudized and expanded in a
plane-wave basis set~\cite{booth2016}. Augmented functions were not
included for the Li atom because they possess a small exponent for the
radial part that introduces linear dependencies in the virtual orbital space.
The AVQZ--$g$ basis set used here does not
encompass $g$ angular momentum functions since the corresponding pseudization
procedure has not yet been implemented in {\sc vasp}. Counterpoise
corrections (CP) to the basis set superposition error
(BSSE)~\cite{boys1970} were included in all correlated quantum-chemical
calculations with plane-waves that employ PGTOs for the virtual states.

Canonical periodic MP2 calculations using PGTOs were performed with the
{\sc vasp} code~\cite{marsman2009,grueneis2010}.  The evaluation of the
two-electron-four-index integrals requires the intermediate
Fourier-transformed overlap densities which are expanded into an auxiliary
plane-wave basis~\cite{marsman2009}. The kinetic energy cutoff $E_{\chi}$ defining this
auxiliary basis set was set to 200~eV. All reported MP2 adsorption energies have
been checked for convergence with respect to this
cutoff. Table~\ref{tab:mp2ecut} shows the convergence of the MP2 adsorption
energy with respect to the cutoff energy.

Periodic CCSD calculations were performed using
the two-electron-four-index integrals
calculated within the PAW method in {\sc vasp}.
To further reduce the computational cost of coupled cluster methods we first
minimize the number of virtual orbitals.
Pseudized Gaussian orbitals were placed only on the top-most layer
of the LiH slab. In a second step, the auxiliary plane-wave basis, required
for the evaluation of the Coulomb integrals employed a kinetic energy cutoff
of 100~eV. MP2 calculations reveal that this approximation yields adsorption energies
that deviate by 3~meV from those obtained using a cutoff of 200~eV as indicated
in Table~\ref{tab:mp2ecut}.

Kats and Manby~\cite{kats2013} proposed an approximation to
CCSD theory that neglects exchange
processes between different clusters which is formally still exact for two electron
systems. The resultant theories have been called distinguishable cluster theories
because they violate the indistinguishability of electrons in a many-electron system.
However, it has been shown that distinguishable cluster approximations such as
distinguishable cluster singles and
doubles (DCSD) correctly dissociate a number of diatomic molecules and
yield very accurate equilibrium geometries and interaction energies for many
molecular systems, outperforming the accuracy of CCSD theory at the same computational
cost~\cite{kats2014,kats2015,kats2016}. Motivated by these findings we also
performed periodic DCSD calculations for the adsorption energy.

Finally, a $\delta$CCSD(T) correction was applied as the difference between
canonical periodic CCSD(T) and MP2 calculations using the AVDZ PGTOs (placed
in the top-most layer) and an H$_2$O+Li$_{8}$H$_{8}$ simulation cell.

\begin{table}
  \caption{\label{tab:mp2ecut} MP2 adsorption energy against the cutoff energy
  $E_{\chi}$ of the auxiliary basis set. One-particle states were expanded in
  a plane-wave basis set with a cutoff of 500~eV, while the virtual states
  were constructed using an AVTZ basis set.}
  \begin{ruledtabular}
    \begin{tabular}{rr}
      \multicolumn{1}{c}{$E_{\chi}$ [eV]}   &  \multicolumn{1}{c}{$E_\textrm{ads}$ [meV]}      \\ \hline
     50      &  $242$   \\
     100     &  $214$   \\
     150     &  $211$   \\
     200     &  $211$   \\
     250     &  $211$   \\
     300     &  $211$   \\
    \end{tabular}
  \end{ruledtabular}
\end{table}

\subsection{Gaussian-basis calculations} \label{sec:computationalcrystal}

The Gaussian-type-orbital-based HF calculations were performed with the {\sc
  crystal} program package~\cite{Crystal14}. To this end a 64-atom
supercell, a $3\times3\times1$ $k$-mesh and tightened integral prescreening
thresholds (TOLINTEG 8 8 8 25 100) were employed. A valence-triple-zeta (VTZ) basis set
combining Ahlrichs' functions for low angular
momentum~\cite{schaefer1992,schaefer1994} and Dunning's cc-pVTZ
basis set for high angular momentum orbitals was used for the H and O
atoms. The Li atoms were described by an optimized basis set already available
from previous calculations on the LiH crystal~\cite{usvyat2011} (basis set A).
The local MP2 (LMP2) and the explicictly correlated local MP2 (LMP2-F12)~\cite{usvyat2013} calculations were performed with
the {\sc cryscor} code. For these calculations, the VTZ basis set was augmented by additional
diffuse orbitals using the dual basis set technique~\cite{usvyat2010} leading
to AVTZ quality. For the O
and H atoms these were the $d$- and $f$- ($p$- and $d$- for H) orbitals from the
aug-cc-pVTZ basis set, for Li: the $s$-, $p$-, $d$- and $f$- orbitals of the basis set
B of Ref.~\onlinecite{usvyat2011}. The effect of the augmented orbitals on the
HF energy was estimated via the first order singles.\cite{usvyat2010}

The correlation
energy was calculated in the direct space, considering H$_2$O--LiH inter-pairs
with inter-orbital separation up to 15~{\AA}.
From 15~{\AA} to infinity the pair-wise R$^{-6}$ extrapolation was employed.\cite{pisani2008}
For the LiH intra-pairs, the (converged) value of 6 {\AA} was used as the inter-orbital cutoff distance.
In the evaluation of the local F12 correction (within the 3*A approximation\cite{werner2007}), which is of much shorter range than LMP2 itself,\cite{usvyat2013} the
pair cutoff distances were reduced to 4 {\AA} and 8 {\AA} for the LiH intra- and
water--LiH inter-pairs, respectively.

The pair-specific truncated virtual spaces of each Wannier function (WF) pair in the projected atomic orbital (PAO)-based
LMP2 is constructed as the union of the two related orbital domains. In our
calculations, the latter comprised for each LiH WF the PAOs on the H atom and
the five nearest neighbour Li atoms. The orbital domains of WF located on water
comprised all three water atoms. The same domains were also employed for the
local resolution of identity (RI) domains~\cite{usvyat2013} in the LMP2-F12 calculations. For the density
fitting of the electron repulsion integrals and the local RI approximation of
the F12 method the auxiliary basis sets of Weigend and
coworkers\cite{weigend02,weigend02b} were used, i.e., aug-cc-pVTZ-mp2fit
and cc-pVTZ-jkfit, respectively.

In the periodic LMP2 and LMP2-F12 calculations the $1s^2$ core states of O and Li were
kept frozen. Nevertheless,
the correlated core contribution of the $1s^2$ states of the Li atoms
was computed at the MP2 level with an aug-cc-pwCVTZ basis set on
the H$_2$O+Li$_{25}$H$_{25}$ cluster using the {\sc molpro} program package~\cite{molpro2010}.
The core-correlation contribution to the interaction is relatively
short-range making further expansion of the cluster not necessary.
Moreover, coupled-cluster calculations on finite
clusters were also performed using the
{\sc molpro} code.

\subsection{DMC calculations} \label{sec:computationaldmc}

DMC calculations have been performed with the {\sc casino} code~\cite{casino},
using Dirac--Fock pseudo-potentials (PP)~\cite{trail05} and trial wave
functions of the Slater--Jastrow type:
\begin{equation}
\Psi_T ( {\bf R} ) = D^\uparrow D^\downarrow e^{J} \; ,
\end{equation}
where $D^\uparrow$ and $D^\downarrow$ are Slater determinants of up- and
down-spin single-electron orbitals, and $e^J$ is the so called Jastrow factor,
which is the exponential of a sum of one-body (electron-nucleus), two-body
(electron-electron) and three body (electron-electron-nucleus) terms. The
parameters in the Jastrow factor were optimised by minimising the variance of
the variational Monte Carlo energy, which for the system with one water
molecule on a two-layer $3 \times 3$ LiH surface supercell was reduced to just
over 1~Ha$^2$ (740~eV$^2$).

Imaginary time evolution of the Schr\"odinger equation has been performed with
the usual short time approximation, using the locality
approximation~\cite{mitas91} to treat the non-local part of the
pseudopotentials.

The single particle orbitals have been obtained by DFT plane-wave calculations
using the local density approximation and a plane-wave cutoff of 3400~eV,
using the {\sc pwscf} package~\cite{pwscf}, and re-expanded in terms of
B-splines~\cite{alfe04}, using the natural B-spline grid spacing given by $ a
= \pi / G_{\rm max} $, where $G_{\rm max}$ is the length of the largest vector
employed in the plane-wave calculations.

The DMC calculations were then performed with no periodic boundary conditions
in the direction perpendicular to the surface, using the Ewald interaction to
model electron-electron interactions.  DMC adsorption energies were computed
as:
\begin{equation}
E_{\rm ads} = E_s - E_b,
\end{equation}
where $E_b$ is the energy of the system with the water molecule at its
equilibrium position on the surface, and $E_s$ that of the system in which the
water molecule has been displaced vertically by 10~\AA, without relaxing its
structure. In the latter
configuration the residual interaction energy between the molecule and the
surface is negligible, and this definition of $E_{\rm ads}$ maximises DMC
cancellation of time step errors~\cite{gillan15,zen16}.

Adsorption energies were calculated using time steps between 0.001 and
0.05~a.u., and we found that with a time step of 0.02~a.u. $E_{\rm ads}$ is
converged to better than 10~meV.

\section{Results} \label{sec:results}

In order to assess the accuracy of different theories and computational procedures,
we study the adsorption of a single water molecule on the (001) surface of LiH.
We present the results of DFT calculations, different periodic MP2 and
coupled-cluster techniques, and compare these methods with DMC. We first
discuss convergence studies of the various theories with respect to the basis
set, finite-size effects, and number of LiH slabs, and then we compare the
adsorption energies of the different methods.

\subsection{Finite-size and basis set convergence}

The finite-size and the basis set convergence studies summarized in this section
employ a 2-layer LiH substrate as shown in Fig.~\ref{fig:adsorption}.

We first discuss the convergence of the DFT-PBE  and HF
adsorption energies with respect to the system size.
DFT-PBE and HF results using different implementations are summarized in
Table~\ref{tab:dft-hf}. Converged
results are in excellent agreement using plane-waves and Gaussian
basis sets, with {\sc vasp} and {\sc crystal} respectively. DFT-PBE results
are converged already with a 32-atom LiH surface slab due to the inability of
DFT-PBE to describe long-range dispersive interactions. HF results
also exhibit a very fast rate of convergence albeit underestimating
the adsorption energy compared to DFT-PBE significantly due to the
neglect of any electronic correlation effects.

\begin{table}
  \caption{\label{tab:dft-hf} DFT-PBE and HF adsorption energies for water on
    2-layer LiH substrates with different number of atoms in the
    supercell and different $k$-meshes. The reference 2-layer
    geometry with 64-atoms is shown in Fig.~\ref{fig:adsorption}.
    The DFT-PBE and HF calculations have been performed with {\sc vasp} and
    employ a 500~eV kinetic energy cutoff. HF {\sc crystal} calculations with
    an AVTZ-quality basis set and a $3\times3\times1$ $k$-mesh yield a value
    of 14~meV.}
  \begin{ruledtabular}
    \begin{tabular}{rrrr}
     \multicolumn{4}{c}{$E_{\textrm{ads}}$ (meV)}  \\
     \multicolumn{1}{c}{$k$-mesh} & \multicolumn{1}{c}{Atoms} & \multicolumn{1}{c}{PBE} & \multicolumn{1}{c}{HF} \\ \hline
     ($\Gamma$-point)     & 32     &    $219$     & $10$ \\
     ($\Gamma$-point)     & 64     &    $215$     & $14$ \\
     ($\Gamma$-point)     & 128    &    $215$     & $15$ \\
     ($3\times3\times1$)  & 64     &    $214$     & $15$ \\
    \end{tabular}
  \end{ruledtabular}
\end{table}

\begin{table}
  \caption{\label{tab:mp2} Canonical MP2 adsorption energies for water on 2-layer LiH substrates
    with different number of atoms in the computational supercell. The calculations were
    performed with {\sc vasp} and employ PGTOs for the virtual orbitals
    alongside the $\Gamma$-point approximation. The thermodynamic limit is
    obtained from an $1/N^{2}$ extrapolation ($N$ denotes the number of atoms
    in the LiH substrate). The LMP2-F12 and LMP2-pF12 adsorption energies are
    238 and 235~meV respectively.}
  \begin{ruledtabular}
    \begin{tabular}{rrrrrr}
     \multicolumn{6}{c}{$E^{\rm MP2}_{\textrm{ads}}$ (meV)} \\ \\
      \multicolumn{1}{c}{Atoms} &  \multicolumn{1}{c}{AVDZ} &
        \multicolumn{1}{c}{AVTZ} & \multicolumn{1}{c}{AVQZ--$g$} &
        \multicolumn{1}{c}{AV(D,T)Z} & \multicolumn{1}{c}{AV(T,Q--$g$)Z} \\ \hline
      32        &  $162$ &  $193$   &   $198$    &   $207$    &   $201$      \\
      64        &  $181$ &  $211$   &   $218$    &   $224$    &   $222$      \\
      72        &  $185$ &  $213$   &   $220$    &   $226$    &   $224$      \\
      128       &  $188$ &  $218$   &   $228$    &   $231$    &   $235$      \\
      $\infty$  &  $189$ &  $219$   &   $227$    &   $231$    &   $233$      \\
    \end{tabular}
  \end{ruledtabular}
\end{table}

\begin{table}
  \caption{\label{tab:dmc} DMC adsorption energies for water on 2-layer LiH substrates
    with different number of atoms in the computational
    supercell~\cite{binnie2011}. The thermodynamic limit is obtained from a
    $1/N^{5/4}$ extrapolation~\cite{drummond2008}.}
  \begin{ruledtabular}
    \begin{tabular}{rr}
      \multicolumn{2}{c}{$E^{\rm DMC}_{\textrm{ads}}$ (meV)} \\
      \multicolumn{1}{c}{Atoms}   &  \multicolumn{1}{c}{CBS}      \\ \hline
      36         &  $167$ (5)     \\
      64         &  $209$ (5)     \\
      100        &  $224$ (8)     \\
      144        &  $239$ (9)     \\
      $\infty$   &  $250$ (7)     \\
    \end{tabular}
  \end{ruledtabular}
\end{table}

We now turn to the discussion of the adsorption energies using
different implementations of MP2 theory.
LMP2-F12 is expected to provide results very close to
the basis set limit, and, with the settings given in 
Sec.~\ref{sec:computationalcrystal}, also very
close to the thermodynamic limit. It
 yields an adsorption energy of 238~meV. The latter value consists of
14 meV of HF, 189 of the frozen-core periodic LMP2/AVTZ, 18 meV of the F12
correction and  17 meV of the core contribution. Using the basis set
correction from the LMP2-pF12 approach, which is an approximation to
LMP2-F12~\cite{grueneis2015}, leads to a similar value of 235~meV.

\begin{figure}
  \centering
  \includegraphics[width=8.6cm]{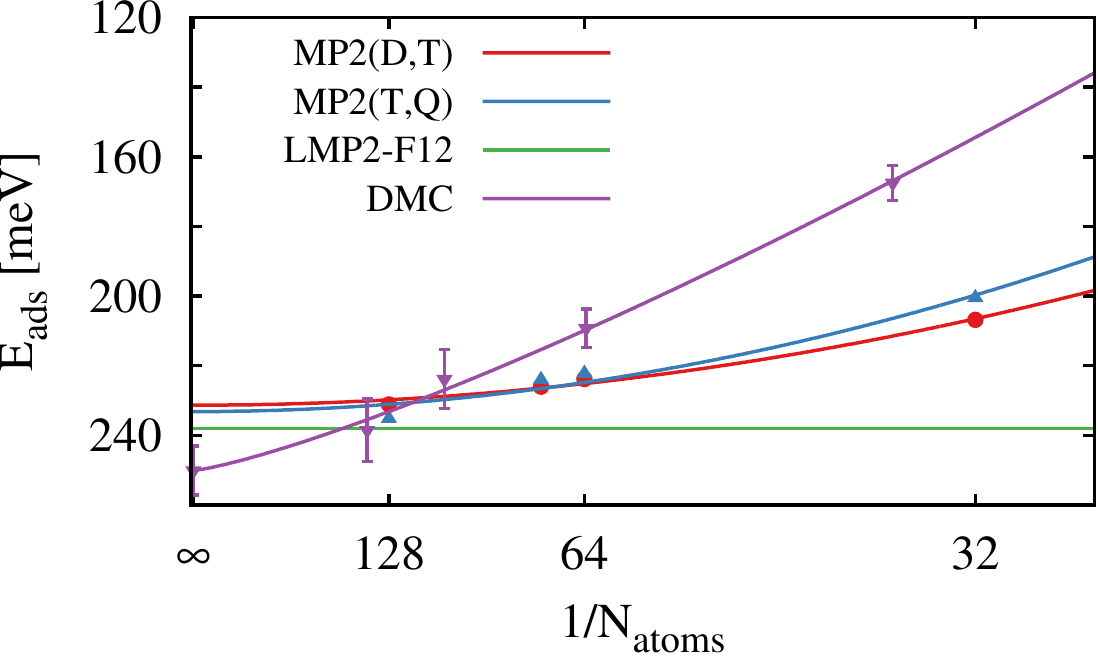}
  \caption{Dependence of the adsorption energy
    $E_{\textrm{ads}}$ of H$_2$O on LiH on the
    number of atoms of the substrate at different levels of theory and basis
    set extrapolations. The fitted lines correspond to $1/N^2$ for the MP2 energies
    and $1/N^{5/4}$ for the DMC energies. MP2 results employ AV(D,T)Z and
    AV(T,Q)Z basis set extrapolations~\cite{halkier}. LMP2-F12 result
    corresponds to the thermodynamic limit. On the $x$-axis
    N$_{\textrm{atoms}}$ is indicated instead of 1$/$N$_{\textrm{atoms}}$.}
  \label{fig:mp2}
\end{figure}

Canonical MP2 energies need to be converged with respect
to both the basis set size and to the LiH surface size.
Table~\ref{tab:mp2} summarizes canonical MP2 adsorption energies obtained
for varying basis set and supercell sizes.
AV(D,T)Z and AV(T,Q--$g$)Z extrapolated adsorption energies agree to within $2-6$~meV for
all studied system sizes. 
We note that the AV(T,Q--$g$)Z extrapolation is somewhat less reliable due to
the absence of $g$ angular momentum functions in the AVQZ values.
We find that the MP2 adsorption energies converge as $1/N^2$,
where $N$ denotes the number of atoms in the LiH substrate.
This behaviour is expected from the long-range decay of
pairwise van der Waals contributions in two-dimensional systems.
The convergence of the finite-size effects for the various basis set extrapolated MP2 results
can be seen in Fig.~\ref{fig:mp2}.
Using the $1/N^2$ behaviour we can extrapolate the MP2 adsorption energies to the thermodynamic limit
($N \rightarrow \infty$), yielding 231~meV and 233~meV for AV(D,T)Z and AV(T,Q--$g$)Z, respectively.
The $5-7$~meV
difference between the canonical MP2 and LMP2-F12 is likely due to the remaining
basis set incompleteness in the correlation energy of the former method.
Notwithstanding, the agreement of the two different schemes,
which have very little in common, is
impressive. The F12-based explicit correlation techniques
combined with local approximation
schemes accelerate the convergence of the MP2 correlation energy. Its close
agreement with the periodic canonical results suggests that PGTOs provide an
adequate virtual basis set for correlated calculations in plane-waves.

DMC adsorption energies~\cite{binnie2011} against the number of atoms in the
simulation supercell are provided in Table~\ref{tab:dmc}.
The DMC adsorption energy converges more slowly with respect
to the supercell size than the MP2 energy as shown in Fig.~\ref{fig:mp2}, due to the
longer ranged nature of the real-space exchange-correlation hole and reduced screening in lower
dimensional materials.
Drummond \textit{et al.}
proposed a $1/N^{5/4}$ extrapolation for two-dimensional
systems~\cite{drummond2008}.
Despite its statistical uncertainty, the thermodynamic limit of the DMC
adsorption energy suggests that the MP2 error for this system is small but not
negligible and thus a higher-order quantum chemical treatment is desirable.

\begin{figure}
  \includegraphics[width=8.6cm]{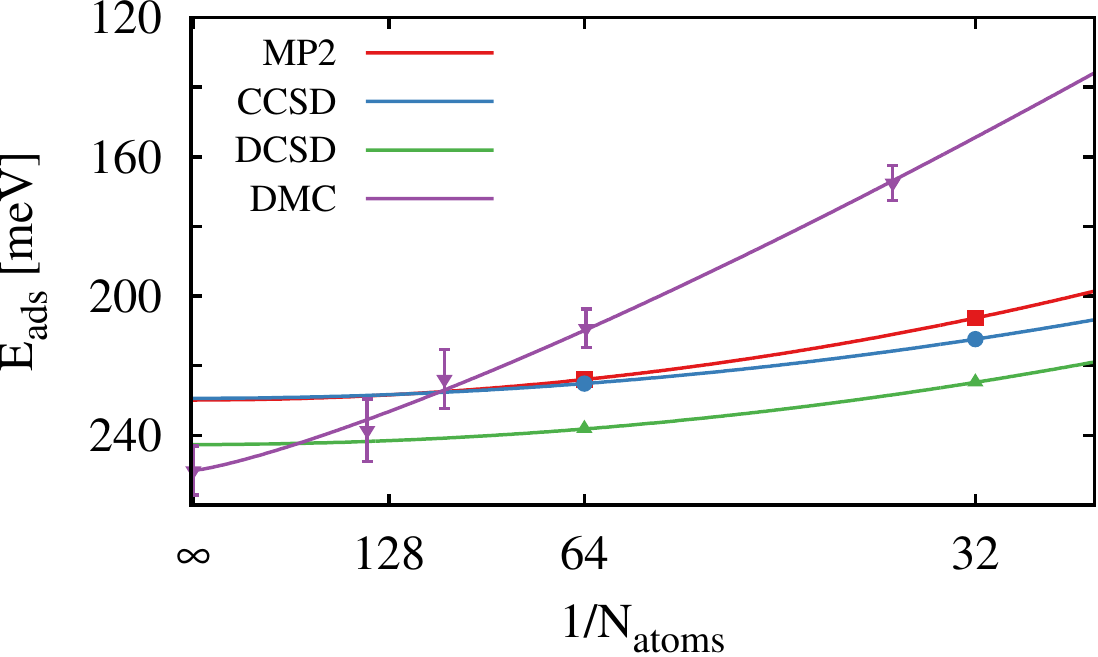}
  \caption{Adsorption energy $E_{\textrm{ads}}$
    of H$_2$O on LiH for different supercells sizes and
    levels of theory. Coupled-cluster and MP2 calculations were done using PGTOs only on the
    top-most layer of the LiH substrate. The fitted lines correspond to
    $1/N^2$ for the coupled-cluster and MP2 energies and $1/N^{5/4}$ for the DMC energies.
    The coupled-cluster and MP2 results employ AV(D,T)Z basis set
    extrapolation~\cite{halkier}. On the $x$-axis
    N$_{\textrm{atoms}}$ is indicated instead of 1$/$N$_{\textrm{atoms}}$.}
  \label{fig:cc-supercell}
\end{figure}

Periodic coupled-cluster calculations were performed with PGTOs for the
virtual orbitals. However, these Gaussian-type functions were placed only on
the top-most layer of the LiH surface to reduce the computational
cost. Additionally, only supercells with 32 and 64 atoms were used to model
the LiH slab.  AVDZ and AVTZ Gaussian basis sets were used for the
construction of the PGTOs, and all results are extrapolated with respect to
the basis set and the number of atoms in the supercell.  MP2 results utilizing
Gaussian orbitals for the full LiH surface and a finite-size extrapolation
using four points, verify that correlation effects are captured adequately
via only top-most layer virtual states and a finite-size extrapolation using
two points.
The error of this simplification is about 1~meV in the MP2 energy.
Consequently, it is reasonable to assume that coupled-cluster results obtained
using the same simplification provide a similarly converged estimate.  MP2 and
coupled-cluster results are summarized in Table~\ref{tab:ccsd} and
Fig.~\ref{fig:cc-supercell}. The CCSD adsorption energies are close to those of
MP2, differing only by 1~meV.  However, the extrapolated DCSD results deviate quite
significantly from the CCSD and MP2 results, yielding an adsorption energy of 243~meV in 
better agreement with the DMC values.

Finally, a $\delta$CCSD(T) correction scheme was applied to both the canonical
and the local MP2 results.
In the former case, the correction $\delta$CCSD(T) was defined as
\begin{equation}
  E^{\delta{\rm CCSD(T)}} =
  E^{{\rm MP2}}_{{\rm H}_2{\rm O+(LiH)}_{\infty}} +
  E^{\rm CCSD(T)}_{{\rm H}_2{\rm O+Li}_8{\rm H}_8} - E^{\rm MP2}_{{\rm H}_2{\rm O+Li}_8{\rm H}_8}
  \ ,\label{eq:delta_cor_pw}
\end{equation}
where canonical CCSD(T) and MP2 calculations were performed using an
H$_2$O+Li$_{8}$H$_{8}$ 2-layer supercell (with an identical orientation of the
water molecule as for the larger supercells) and an AVDZ basis set in a plane-wave
representation. $E^{{\rm MP2}}_{{\rm H}_2{\rm O+(LiH)}_{\infty}}$ is
the thermodynamic limit of the MP2 adsorption energy using
AVD(T,Q--$g$)Z basis set extrapolation. This yields an adsorption
energy of 254~meV.

The $\delta$CCSD(T) corrections to the LMP2-F12 results were computed using finite clusters.
In this case the canonical MP2 and CCSD(T) adsorption energy
calculations were done on an H$_2$O+Li$_{9}$H$_{9}$ 2-layer finite cluster
using the AV(D,T)Z basis sets. The water molecule geometry was taken from the
periodic supercells. The correction $\delta$CCSD(T)
for the periodic system was defined as
\begin{equation}
  E^{\delta{\rm CCSD(T)}} =
  E^{{\rm LMP2-F12}}_{{\rm H}_2{\rm O+(LiH)}_{\infty}} +
  E^{\rm CCSD(T)}_{{\rm H}_2{\rm O+Li}_9{\rm H}_9} - E^{\rm MP2}_{{\rm H}_2{\rm O+Li}_9{\rm H}_9} \ ,\label{eq:delta_cor}
\end{equation}
yielding an adsorption energy of 256~meV.
Incidentaly we note that one cannot construct a periodic Li$_9$H$_9$
supercell and therefore a Li$_8$H$_8$ slab was used for the plane-wave
based $\delta$CCSD(T).
Furthermore, the finite-size error of the correction was estimated
as the difference between local LCCSD(T0)$|$LCCD[S]-R$^{-6}$ calculations\cite{werner2011,masur2013,schutz2014}
on H$_2$O+Li$_{9}$H$_{9}$ and H$_2$O+Li$_{25}$H$_{25}$ clusters. This
difference turned out to be of the order of 0.3~meV. However, we note that a
$\delta$CCSD correction, defined in a analogous way as $\delta$CCSD(T), provides
 an adsorption energy of 219~meV, which deviates somewhat from the periodic CCSD
result. In contrast, a periodic $\delta$CCSD correction, defined in a
analogous way as $\delta$CCSD(T), yields an adsorption energy of 227~meV, very
close to the canonical CCSD result.
Thus the finite-cluster $\delta$ approach might still
contain a certain error.

\begin{table}
  \caption{\label{tab:ccsd} MP2 and coupled-cluster adsorption energies using
    LiH substrates with different number of atoms in the supercell. PGTOs were
    used for the virtual orbitals in the top-most layer of the LiH surface.
    The thermodynamic limit is obtained via an $1/N^2$ extrapolation.}
  \begin{ruledtabular}
    \begin{tabular}{rrrr}
      \multicolumn{4}{c}{$E^{\rm MP2}_{\textrm{ads}}$ (meV)} \\ \\
      \multicolumn{1}{c}{Atoms} & \multicolumn{1}{c}{AVDZ} &
        \multicolumn{1}{c}{AVTZ} & \multicolumn{1}{c}{AV(D,T)Z} \\ \hline
      32        &    $157$   & $192$         &   $207$          \\
      64        &    $173$   & $209$         &   $224$          \\
      $\infty$  &    $180$   & $216$         &   $230$          \\ \\
      \multicolumn{4}{c}{$E^{\rm CCSD}_{\textrm{ads}}$ (meV)} \\ \\
      Atoms & AVDZ   & AVTZ  & AV(D,T)Z \\ \hline
      32        &    $152$   & $195$         &   $212$          \\
      64        &    $172$   & $209$         &   $225$          \\
      $\infty$  &    $180$   & $215$         &   $229$          \\ \\
      \multicolumn{4}{c}{$E^{\rm DCSD}_{\textrm{ads}}$ (meV)} \\ \\
      Atoms & AVDZ   & AVTZ  & AV(D,T)Z \\ \hline
      32        &    $162$   & $206$         &   $225$          \\
      64        &    $183$   & $222$         &   $238$          \\
      $\infty$  &    $192$   & $229$         &   $243$          \\
    \end{tabular}
  \end{ruledtabular}
\end{table}

\subsection{Convergence of the adsorption energy with the number of layers}\label{sec:layers}

In this section we investigate the adequacy of the chosen slab model, which
consists of just two LiH layers, for studying adsorption of water. Generally,
the convergence of the adsorption energy with the number of layers in the slab
is expected to be governed by long-range effects, such as electrostatics
(attractive or repulsive) and dispersion (attractive). Importantly,
electrostatics are already captured at the DFT or HF levels, while dispersion
is not (unless the dispersion correction is added or a special DFT functional
is used, that is able to describe dispersion).

Table~\ref{tab:layers} demonstrates by how much the adsorption energy grows or
declines if further layers are added to the slab, as computed by DFT and
HF. In order to isolate the dispersion contribution, we provide the -D3
contribution separately, as well as the LMP2 correlation energy. For
dispersion alone it is actually possible to obtain convergence with the number
of layers: -D3 is very inexpensive and thus can be computed for very thick
slabs, while for LMP2 the inter-adsorbate-slab contribution can be
extrapolated to a semi-infinite slab using the pair-specific $C_6$
coefficients fitted to the actual LMP2 pair energies (see
Ref.~\onlinecite{pisani2012} for details).

The PBE and HF results suggest that for the non-dispersive contributions, the
two-layer slab is already an adequate model. Dispersion on the contrary, is
not entirely converged with just two LiH-layers. However, at the scale of the
whole adsorption energy, the lack of a few meV of dispersion in the two-layer
model can be tolerated.


\begin{table}
\caption{\label{tab:layers} Convergence of the
      adsorption energy (DFT-PBE, HF), the dispersion correction (-D3), and
      the correlation energy (LMP2) with respect to the slab thickness. The
      provided energies (in meV) represent the
      excess or depletion in the energy with respect to the 2-layer
      slab model due to additional layers. All the calculations  employed the
      $4\times4$ surface supercell. The $\infty$ symbol
      indicates the converged D3 and LMP2 value. The latter is obtained by
      extrapolation of the inter-LiH-Water energy from the 3-layer model to a
      semi-infinite slab by means of the slab replication technique of
      Ref. \onlinecite{pisani2012}, employing pair-specific $C_6$ coefficients fitted to
      the actual LMP2 pair energies. The result of such an extrapolation from the
      2-layer model is given in the parenthesis.}
 \begin{ruledtabular}
    \begin{tabular}{lrrrl}
      \multicolumn{1}{c}{No. of layers}   & \multicolumn{1}{c}{PBE} & \multicolumn{1}{c}{HF} & \multicolumn{1}{c}{-D3} & \multicolumn{1}{c}{LMP2} \\ \hline
      3           &  $-0.15$  & $-1.51$ & $+5.36$ & $+2.44$         \\
      4           &  $-0.16$  &         & $+7.01$ &                 \\
      $\infty$    &           &         & $+8.44$ & $+4.66 (+4.97)$ \\
    \end{tabular}
  \end{ruledtabular}
\end{table}

\begin{figure*}
  \includegraphics[width=17.8cm]{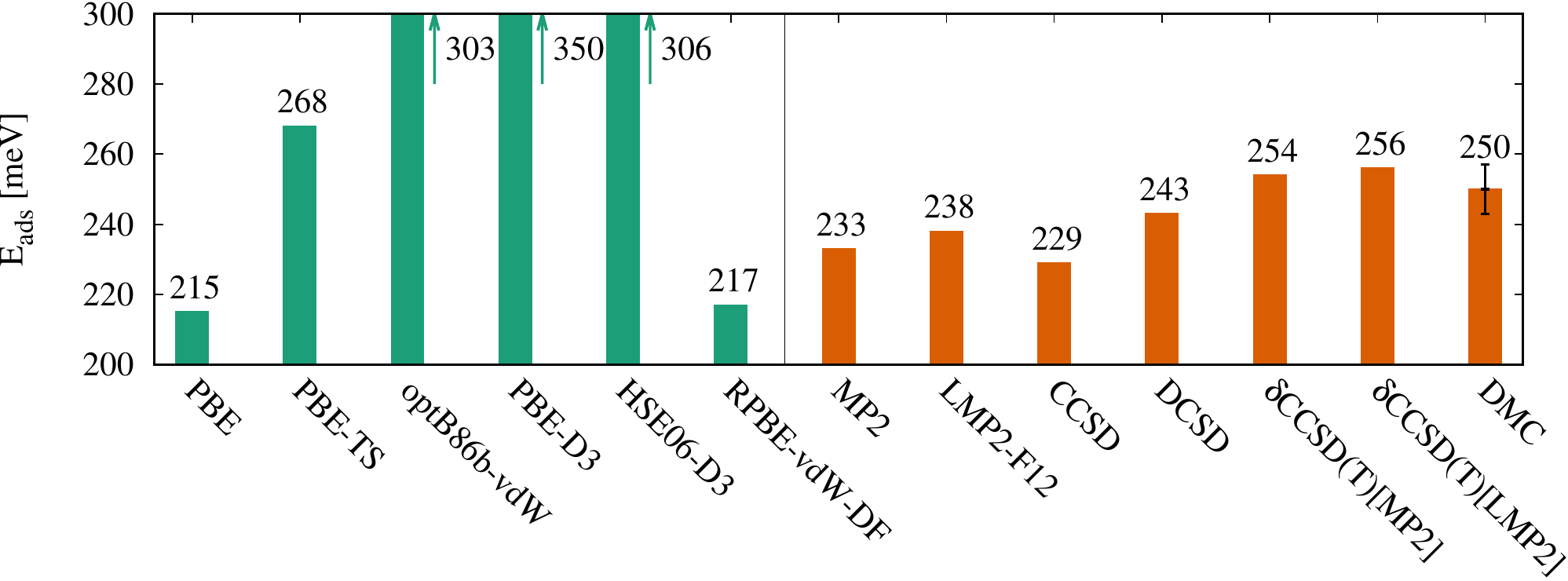}
  \caption{Converged adsorption energies of a water molecule on a LiH surface at different
    levels of theory. PBE and several van der Waals functionals
    shown on the left. Wave function based methods ranging from MP2 to
    $\delta$CCSD(T) and DMC shown on the right.}
  \label{fig:comparison}
\end{figure*}

\subsection{Comparison of methods}

We now summarize the converged adsorption energies and compare them to a small
set of widely-used density-functionals.  All reported results employ a 2-layer
LiH substrate as
in Fig.~\ref{fig:adsorption}.  We believe that the mutually agreeing DMC and
$\delta$CCSD(T) results can be considered as the most reliable benchmark for
the present system, yielding adsorption energies between 250 ($\pm7$)~meV and
256~meV.  For comparison, the adsorption energy of each method is depicted in
Fig.~\ref{fig:comparison}.  A sizeable variation in the adsorption energies is
evident between different van der Waals functionals
(PBE-TS~\cite{tkatchenko2009}, optB86b-vdW~\cite{klimes2011},
PBE-D3~\cite{grimme2010}, HSE06-D3~\cite{moellman2014},
RPBE-vdW-DF~\cite{dion2004}), as well as PBE.  The PBE functional
underestimates the adsorption energy by roughly 45~meV, in a large part due to
its lack of dispersive interactions. Grimme's D3 correction~\cite{grimme2010}
accounts for such interactions, albeit overestimating the adsorption energy
for the current system, predicting a PBE-D3 adsorption energy of 350~meV,
consistent with similar findings for water adsorption on ionic surfaces~\cite{kanaki2015}.
We note that this overestimation is less pronounced
when the HSE06~\cite{heyd2003,krukau2006} hybrid functional is used in conjunction with D3,
yielding a value of 306~meV.
This can partly be attributed to the fact that 
the HSE06 functional underestimates the adsorption energy
compared to PBE by as much as 85~meV.
The optB86b-vdW~\cite{klimes2011} results
also overbind the water molecule by roughly~45~meV, while the
RPBE-vdW-DF~\cite{dion2004} adsorption energy exhibits a similar underbinding
as for the case of PBE. The best van der Waals functional estimate is
provided by the Tkatchenko and Scheffler functional (PBE-TS) with iterative
Hirshfeld partitioning~\cite{bucko2013,bucko2014}. The latter yields an
adsorption energy of~268~meV in good agreement with $\delta$CCSD(T) results.
These results illustrate the difficulties in van der Waals functionals.
The PBE functional is known to provide a non-electrostatic binding
between closed shell systems. This attraction
is rather an artifact than a real dispersive interaction. At the same time,
this artificial attraction provides a quantitatively reasonable effective
substitute for dispersion. However, if the physically correct dispersion is
added on top, it becomes difficult to avoid double counting, leading to a
deterioration of the quantitative accuracy.

Figure~\ref{fig:comparison} also shows the various wave function estimates of
the adsorption energy.  Canonical MP2 theory underestimates the adsorption energy by
17~meV compared to DMC, while LMP2-F12 provides a slightly better estimate,
partly due to the explicit correlation, leading to an improved convergence with
respect to the basis set size.  The LMP2-F12 adsorption energy is 238~meV,
only 11~meV below the DMC result.  CCSD constitutes no improvement over MP2
theory for the present case, yielding a binding energy of 229~meV only.
The DCSD approximation~\cite{kats2013}, consistent with findings in
molecular systems~\cite{kats2013,kats2014,kats2015}, considerably improves the
description of water adsorption on LiH, predicting an adsorption energy of
243~meV, which is within the stochastic error of DMC but still underbinding
compared to the triples corrected $\delta$CCSD(T) results.  In summary, we
find excellent agreement between high-level quantum chemistry and QMC
techniques as well as between different methods to compute MP2 adsorption
energies. Furthermore the correlated wave function based methods yield
estimates for the binding energy that lie in a relatively narrow energy window
ranging from 229~meV to 256~meV.

\section{Conclusions}  \label{sec:conclusions}

We have presented a comprehensive comparison between different electronic
structure methods including wave function based theories and a small selection
of density-functionals for the prediction of the adsorption
energy of a single water molecule on the (001) LiH surface.

Quantum chemical methods are systematically improvable, hence yielding
increasingly accurate adsorption energies as one moves up the hierarchy to
higher orders of theory. Distinguishable cluster theory and inclusion of
triple excitations to CCSD theory give the best agreement with DMC results.
We find that MP2 and CCSD reach a similar level of accuracy for this system,
slightly underbinding the water molecule on the LiH surface by roughly 20~meV.
We also find good agreement between periodic canonical and local
implementations utilizing explicit correlation technique for improved basis
set convergence. All this demonstrates that quantum chemical approaches are
becoming a robust and reliable tool for condensed phase electronic structure
calculations.

We have also employed van der Waals functionals for the study of the same
system, finding that these functionals yield a significantly larger spread of
adsorption energy estimates compared to the employed many-electron theories.
The under- and overestimations compared to DMC and $\delta$CCSD(T) are as
large as 30~meV (RPBE-vdW-DF) and 100~meV (PBE-D3), respectively.  Although
the PBE-TS functional achieves good agreement with the DMC and
$\delta$CCSD(T) estimates for the present case, it remains difficult to
achieve such a high level of accuracy for a wide class of materials using
van der Waals functionals. This study contributes another benchmark
system to the literature that can be used to further improve upon the
currently available and computationally very efficient van der Waals
functionals for cases where higher accuracy is needed.

\begin{acknowledgements}
AM's work is supported by the European Research Council under the European
Union's Seventh Framework Programme (No. FP/2007-2013)/ERC Grant Agreement
No.~616121 (HeteroIce project) and the Royal Society through a Wolfson
Research Merit Award. DU and MS acknowledge financial support from the Deutsche
Forschungsgemeinschaft (Grants US-103/1-2 and SCHU-1456/12-1). GHB is grateful
for support from the Royal Society via a University Research Fellowship.
\end{acknowledgements}

\bibliographystyle{apsrev4-1}
\bibliography{bibliography}

\newpage
  
\section*{Supplementary Material}

We have included the structural coordinates of four different adsorption
geomerties corresponding to 16-, 32-, 64-, and 128-atoms LiH two-layer
substrates in a {\sc poscar} format for {\sc vasp} calculations.

\begin{verbatim}
H2O-Li8H8
  1.0000000000
     5.7756481171  0.0000000000  0.0000000000
     0.0000000000  5.7756481171  0.0000000000
     0.0000000000  0.0000000000 20.4200000763
  Li H O
   8 8 1
Cartesian coordinates
   0.000000000  0.000000000  0.000000000
   0.000000000  2.887823886  0.000000000
   2.887823886  0.000000000  0.000000000
   2.887823886  2.887823886  0.000000000
   1.443911943  1.443911943  2.041999886
   1.443911943  4.331735744  2.041999886
   4.331735744  1.443911943  2.041999886
   4.331735744  4.331735744  2.041999886
   1.443911943  1.443911943  0.000000000
   1.443911943  4.331735744  0.000000000
   4.331735744  1.443911943  0.000000000
   4.331735744  4.331735744  0.000000000
   0.000000000  0.000000000  2.041999886
   0.000000000  2.887823886  2.041999886
   2.887823886  0.000000000  2.041999886
   2.887823886  2.887823886  2.041999886
   2.127302827  0.674412056  4.089802120
   2.127302827  2.213411788  4.089802120
   1.532717715  1.443911943  4.193270986
\end{verbatim}

\newpage

\begin{verbatim}
H2O-Li16H16
  1.0000000000
     8.168000000  0.000000000  0.000000000
     0.000000000  8.168000000  0.000000000
     0.000000000  0.000000000 20.419999999
  Li  H O
  16 18 1
Cartesian coordinates
   0.00000000  0.00000000  0.00000000
   0.00000000  4.08400000  0.00000000
   4.08400000  0.00000000  0.00000000
   4.08400000  4.08400000  0.00000000
   0.00000000  2.04200000  2.04200000
   0.00000000  6.12600000  2.04200000
   4.08400000  2.04200000  2.04200000
   4.08400000  6.12600000  2.04200000
   2.04200000  0.00000000  2.04200000
   2.04200000  4.08400000  2.04200000
   6.12600000  0.00000000  2.04200000
   6.12600000  4.08400000  2.04200000
   2.04200000  2.04200000  0.00000000
   2.04200000  6.12600000  0.00000000
   6.12600000  2.04200000  0.00000000
   6.12600000  6.12600000  0.00000000
   2.04200000  2.04200000  2.04200000
   2.04200000  6.12600000  2.04200000
   6.12600000  2.04200000  2.04200000
   6.12600000  6.12600000  2.04200000
   2.04200000  0.00000000  0.00000000
   2.04200000  4.08400000  0.00000000
   6.12600000  0.00000000  0.00000000
   6.12600000  4.08400000  0.00000000
   0.00000000  2.04200000  0.00000000
   0.00000000  6.12600000  0.00000000
   4.08400000  2.04200000  0.00000000
   4.08400000  6.12600000  0.00000000
   0.00000000  0.00000000  2.04200000
   0.00000000  4.08400000  2.04200000
   4.08400000  0.00000000  2.04200000
   4.08400000  4.08400000  2.04200000
   3.06934903  4.02311175  4.08980207
   1.98111175  5.11134903  4.08980207
   2.10479519  4.14679519  4.19327115
\end{verbatim}

\newpage

\begin{verbatim}
H2O-Li32H32
  1.0000000000
     8.168000000  8.168000000  0.000000000
    -8.168000000  8.168000000  0.000000000
     0.000000000  0.000000000 20.420000000
  Li H  O
  32 34 1
Cartesian coordinates
   0.00000000  0.00000000  0.00000000
  -2.04200000  2.04200000  0.00000000
  -4.08400000  4.08400000  0.00000000
  -6.12600000  6.12600000  0.00000000
   2.04200000  2.04200000  0.00000000
   0.00000000  4.08400000  0.00000000
  -2.04200000  6.12600000  0.00000000
  -4.08400000  8.16800000  0.00000000
   4.08400000  4.08400000  0.00000000
   2.04200000  6.12600000  0.00000000
   0.00000000  8.16800000  0.00000000
  -2.04200000 10.21000000  0.00000000
   6.12600000  6.12600000  0.00000000
   4.08400000  8.16800000  0.00000000
   2.04200000 10.21000000  0.00000000
   0.00000000 12.25200000  0.00000000
   0.00000000  2.04200000  2.04200000
  -2.04200000  4.08400000  2.04200000
  -4.08400000  6.12600000  2.04200000
  -6.12600000  8.16800000  2.04200000
   2.04200000  4.08400000  2.04200000
   0.00000000  6.12600000  2.04200000
  -2.04200000  8.16800000  2.04200000
  -4.08400000 10.21000000  2.04200000
   4.08400000  6.12600000  2.04200000
   2.04200000  8.16800000  2.04200000
   0.00000000 10.21000000  2.04200000
  -2.04200000 12.25200000  2.04200000
   6.12600000  8.16800000  2.04200000
   4.08400000 10.21000000  2.04200000
   2.04200000 12.25200000  2.04200000
   0.00000000 14.29400000  2.04200000
   0.00000000  2.04200000  0.00000000
  -2.04200000  4.08400000  0.00000000
  -4.08400000  6.12600000  0.00000000
  -6.12600000  8.16800000  0.00000000
   2.04200000  4.08400000  0.00000000
   0.00000000  6.12600000  0.00000000
  -2.04200000  8.16800000  0.00000000
  -4.08400000 10.21000000  0.00000000
   4.08400000  6.12600000  0.00000000
   2.04200000  8.16800000  0.00000000
   0.00000000 10.21000000  0.00000000
  -2.04200000 12.25200000  0.00000000
   6.12600000  8.16800000  0.00000000
   4.08400000 10.21000000  0.00000000
   2.04200000 12.25200000  0.00000000
   0.00000000 14.29400000  0.00000000
   0.00000000  0.00000000  2.04200000
  -2.04200000  2.04200000  2.04200000
  -4.08400000  4.08400000  2.04200000
  -6.12600000  6.12600000  2.04200000
   2.04200000  2.04200000  2.04200000
   0.00000000  4.08400000  2.04200000
  -2.04200000  6.12600000  2.04200000
  -4.08400000  8.16800000  2.04200000
   4.08400000  4.08400000  2.04200000
   2.04200000  6.12600000  2.04200000
   0.00000000  8.16800000  2.04200000
  -2.04200000 10.21000000  2.04200000
   6.12600000  6.12600000  2.04200000
   4.08400000  8.16800000  2.04200000
   2.04200000 10.21000000  2.04200000
   0.00000000 12.25200000  2.04200000
   1.02734903  1.98111175  4.08980207
  -0.06088825  3.06934903  4.08980207
   0.06279519  2.10479519  4.19327115
\end{verbatim}

\newpage

\begin{verbatim}
H2O-Li64H64
  1.000000000
    16.336000000  0.000000000  0.000000000
     0.000000000 16.336000000  0.000000000
     0.000000000  0.000000000 20.000000000
  Li  H  O 
  64  66 1
Cartesian coordinates
   0.00000000  0.00000000  0.00000000
   0.00000000  4.08400000  0.00000000
   0.00000000  8.16800000  0.00000000
   0.00000000 12.25200000  0.00000000
   4.08400000  0.00000000  0.00000000
   4.08400000  4.08400000  0.00000000
   4.08400000  8.16800000  0.00000000
   4.08400000 12.25200000  0.00000000
   8.16800000  0.00000000  0.00000000
   8.16800000  4.08400000  0.00000000
   8.16800000  8.16800000  0.00000000
   8.16800000 12.25200000  0.00000000
  12.25200000  0.00000000  0.00000000
  12.25200000  4.08400000  0.00000000
  12.25200000  8.16800000  0.00000000
  12.25200000 12.25200000  0.00000000
   0.00000000  2.04200000  2.04200000
   0.00000000  6.12600000  2.04200000
   0.00000000 10.21000000  2.04200000
   0.00000000 14.29400000  2.04200000
   4.08400000  2.04200000  2.04200000
   4.08400000  6.12600000  2.04200000
   4.08400000 10.21000000  2.04200000
   4.08400000 14.29400000  2.04200000
   8.16800000  2.04200000  2.04200000
   8.16800000  6.12600000  2.04200000
   8.16800000 10.21000000  2.04200000
   8.16800000 14.29400000  2.04200000
  12.25200000  2.04200000  2.04200000
  12.25200000  6.12600000  2.04200000
  12.25200000 10.21000000  2.04200000
  12.25200000 14.29400000  2.04200000
   2.04200000  0.00000000  2.04200000
   2.04200000  4.08400000  2.04200000
   2.04200000  8.16800000  2.04200000
   2.04200000 12.25200000  2.04200000
   6.12600000  0.00000000  2.04200000
   6.12600000  4.08400000  2.04200000
   6.12600000  8.16800000  2.04200000
   6.12600000 12.25200000  2.04200000
  10.21000000  0.00000000  2.04200000
  10.21000000  4.08400000  2.04200000
  10.21000000  8.16800000  2.04200000
  10.21000000 12.25200000  2.04200000
  14.29400000  0.00000000  2.04200000
  14.29400000  4.08400000  2.04200000
  14.29400000  8.16800000  2.04200000
  14.29400000 12.25200000  2.04200000
   2.04200000  2.04200000  0.00000000
   2.04200000  6.12600000  0.00000000
   2.04200000 10.21000000  0.00000000
   2.04200000 14.29400000  0.00000000
   6.12600000  2.04200000  0.00000000
   6.12600000  6.12600000  0.00000000
   6.12600000 10.21000000  0.00000000
   6.12600000 14.29400000  0.00000000
  10.21000000  2.04200000  0.00000000
  10.21000000  6.12600000  0.00000000
  10.21000000 10.21000000  0.00000000
  10.21000000 14.29400000  0.00000000
  14.29400000  2.04200000  0.00000000
  14.29400000  6.12600000  0.00000000
  14.29400000 10.21000000  0.00000000
  14.29400000 14.29400000  0.00000000
   2.04200000  2.04200000  2.04200000
   2.04200000  6.12600000  2.04200000
   2.04200000 10.21000000  2.04200000
   2.04200000 14.29400000  2.04200000
   6.12600000  2.04200000  2.04200000
   6.12600000  6.12600000  2.04200000
   6.12600000 10.21000000  2.04200000
   6.12600000 14.29400000  2.04200000
  10.21000000  2.04200000  2.04200000
  10.21000000  6.12600000  2.04200000
  10.21000000 10.21000000  2.04200000
  10.21000000 14.29400000  2.04200000
  14.29400000  2.04200000  2.04200000
  14.29400000  6.12600000  2.04200000
  14.29400000 10.21000000  2.04200000
  14.29400000 14.29400000  2.04200000
   2.04200000  0.00000000  0.00000000
   2.04200000  4.08400000  0.00000000
   2.04200000  8.16800000  0.00000000
   2.04200000 12.25200000  0.00000000
   6.12600000  0.00000000  0.00000000
   6.12600000  4.08400000  0.00000000
   6.12600000  8.16800000  0.00000000
   6.12600000 12.25200000  0.00000000
  10.21000000  0.00000000  0.00000000
  10.21000000  4.08400000  0.00000000
  10.21000000  8.16800000  0.00000000
  10.21000000 12.25200000  0.00000000
  14.29400000  0.00000000  0.00000000
  14.29400000  4.08400000  0.00000000
  14.29400000  8.16800000  0.00000000
  14.29400000 12.25200000  0.00000000
   0.00000000  2.04200000  0.00000000
   0.00000000  6.12600000  0.00000000
   0.00000000 10.21000000  0.00000000
   0.00000000 14.29400000  0.00000000
   4.08400000  2.04200000  0.00000000
   4.08400000  6.12600000  0.00000000
   4.08400000 10.21000000  0.00000000
   4.08400000 14.29400000  0.00000000
   8.16800000  2.04200000  0.00000000
   8.16800000  6.12600000  0.00000000
   8.16800000 10.21000000  0.00000000
   8.16800000 14.29400000  0.00000000
  12.25200000  2.04200000  0.00000000
  12.25200000  6.12600000  0.00000000
  12.25200000 10.21000000  0.00000000
  12.25200000 14.29400000  0.00000000
   0.00000000  0.00000000  2.04200000
   0.00000000  4.08400000  2.04200000
   0.00000000  8.16800000  2.04200000
   0.00000000 12.25200000  2.04200000
   4.08400000  0.00000000  2.04200000
   4.08400000  4.08400000  2.04200000
   4.08400000  8.16800000  2.04200000
   4.08400000 12.25200000  2.04200000
   8.16800000  0.00000000  2.04200000
   8.16800000  4.08400000  2.04200000
   8.16800000  8.16800000  2.04200000
   8.16800000 12.25200000  2.04200000
  12.25200000  0.00000000  2.04200000
  12.25200000  4.08400000  2.04200000
  12.25200000  8.16800000  2.04200000
  12.25200000 12.25200000  2.04200000
   3.06934903  4.02311175  4.08980207
   1.98111175  5.11134903  4.08980207
   2.10479519  4.14679519  4.19327115
\end{verbatim}

\end{document}